\documentstyle[12pt,epsfig]{article}

\newcommand{\Etmax}{\mbox{$E_{T}^{\rm max}$}}
\begin{document}

%%%%%%%%%%%%%% title page contents
\begin{titlepage}
\def\thepage {}     % kill page numbering

\title{Dijet Mass Spectrum Limits on \\ Flavor-Universal Colorons }
\author{Iain Bertram$^{a*}$ and Elizabeth H. Simmons$^b$\thanks{e-mail
    addresses:
bertram@fnal.gov, simmons@bu.edu} \\
${}^a$ Department of Physics, Northwestern University \\
2145 Sheridan Road, Evanston, IL  60208 \\
${}^b$ Department of Physics, Boston University \\
590 Commonwealth Ave., Boston  MA  02215}

\date{\today}
\maketitle

\bigskip
\begin{picture}(0,0)(0,0)
\put(295,280){BUHEP-98-24}
\put(295,265){hep-ph/9809472}
\end{picture}
\vspace{24pt}

\begin{abstract}
  Using recent D\O\ data on the dijet mass spectrum, we present a
  limit on flavor-universal colorons.  At 95\% CL we find
  $M_c/\cot\theta > 837$ GeV.  We discuss the implications of
  this limit for models of quark compositeness, non-standard gluon
  interactions, and dynamical electroweak symmetry breaking.  In
  addition, we place a lower bound $\Lambda_{A8} > 2.1$ TeV on the scale
  of color-octet axial-vector contact interactions among quarks which
  could arise in models of quark compositeness.
\pagestyle{empty}
\end{abstract}

\end{titlepage}

%%%%%%%%%%%%%%

%%%%%%%%%%%%%%
\section{Introduction}
\label{sec:intro}
\setcounter{equation}{0}

The flavor-universal coloron model \cite{newint} was originally proposed
to explain the apparent excess of high-$E_T$ jets in the inclusive jet
spectrum measured at the Fermilab Tevatron by the CDF Collaboration
\cite{CDFexc}.  This model is a flavor-universal variant of the coloron
model of Hill and Parke \cite{topglu} which involves a minimal extension
of the standard description of the strong interactions, including the
addition of one gauge interaction and a scalar multiplet, but no new
fermions.  The flavor-universal coloron model of the strong interactions
can be grafted onto the standard one-Higgs-doublet model of electroweak
physics, yielding a simple, complete, and renormalizable theory.
Alternatively, it can provide the basis for dynamical generation of
electroweak symmetry breaking and the generation of the top quark's mass
in models \cite{zpcol,kdlr} akin to top-color-assisted technicolor
\cite{tc2}.

Previous work on the phenomenology of the colorons has considered
effects on the $\rho$ parameter \cite{newint,zpcol}, the inclusive jet
spectrum \cite{newint}, the dijet spectrum and angular distributions
\cite{rmh,colphen,anglim}, and b-tagged dijets \cite{colphen}.  The most
recent of these analyses \cite{anglim} put a limit $M_c/\cot\theta >
759$ GeV on the the coefficient of the four-fermion contact interaction
to which heavy coloron exchange would give rise.

This letter explores the effects colorons would have on the dijet mass
spectrum measured at the Tevatron by the D\O\ Collaboration
\cite{didata} and establishes a still stronger limit on contact
interactions arising from colorons: $M_c/\cot\theta > 837$ GeV at 95\%
CL.  In Section 2, we briefly review the model.  Section 3 explains how
our limit was derived.  Section 4 discusses our conclusions and the
implications for models of quark compositeness, non-standard gluon
interactions, and dynamical electroweak symmetry breaking.  We also
present a separate limit on the scale of color-octet axial-vector contact
interactions among quarks.

%%%%%%%%%%%%%%
\section{The model}
\label{sec:model}
\setcounter{equation}{0}

In the flavor-universal coloron model  \cite{newint}, the strong gauge
group is extended to $SU(3)_1 \times SU(3)_2$.  The gauge couplings
are, respectively, $\xi_1$ and $\xi_2$ with $\xi_1 \ll \xi_2$.  Each
quark transforms as a (1,3) under this extended strong gauge group.

The model also includes a scalar boson $\Phi$ transforming as a $(3,\bar
3)$ under the two $SU(3)$ groups.  For a range of couplings in the
scalar potential, $\Phi$ develops a vacuum expectation value
$\langle\Phi\rangle = {\rm diag}(f,f,f)$ which breaks the two strong
groups to their diagonal subgroup \cite{newint}.  We identify this
unbroken subgroup with QCD.

When the extended color symmetry breaks, the original gauge bosons mix
to form an octet of massless gluons and an octet of massive colorons.
The gluons interact with quarks through a conventional QCD coupling with
strength $g_3$.  The colorons $(C^{\mu a})$ interact with quarks through
a new QCD-like coupling
\begin{equation}
{\cal L} = - g_3  \cot\theta J^a_\mu C^{\mu a} \ \ ,
\end{equation}
where  $J^a_\mu$ is the color current
\begin{equation}
\sum_f {\bar q}_f \gamma_\mu \frac{\lambda^a}{2}q_f \ \ .
\end{equation} 
and $\cot\theta = \xi_2/\xi_1\ > 1$.  The mass of the colorons may be
written 
\begin{equation}
M_C = \left({g_3 \over \sin\theta \cos\theta}\right) f
\end{equation}
in terms of the parameters of the model.

Below the scale $M_C$, coloron-exchange may be approximated
by the effective four-fermion interaction
\begin{equation}
{\cal L}_{eff} = -
{g_3^2\cot^2\theta\over {2! M_C^2}} J_\mu^a J^{\mu a}\ .
\label{conta}
\end{equation}
This can be re-written in the form commonly used in studies of
quark compositeness \cite{elp}
\begin{equation}
{\cal L}_{eff} = -
{4\pi \over {2! \Lambda_{V8}^2}} J_\mu^a J^{\mu a}\ .
\label{lamta}
\end{equation}
where the scale $\Lambda_{V8}$ is defined by $\Lambda_{V8} \sqrt{\alpha_s} =
M_c/\cot\theta$.  Contact interactions of this kind tend to increase
quark-quark scattering at high invariant mass above the standard QCD
prediction.

%%%%%%%%%%%%%%
\section{Effects on the Dijet Spectrum}
\label{sec:spectrum}
\setcounter{equation}{0}

The D\O\ Collaboration recently \cite{didata} measured the
inclusive dijet mass spectrum at $\sqrt{s}$= 1.8 TeV for dijet masses
above 200 GeV and jet pseudorapidity $\vert\eta_{jet}\vert < 1.0$.  The
collaboration also measured the ratio of spectra at $\vert\eta_{jet}\vert
< 0.5$ and $0.5<\vert\eta_{jet}\vert<1.0$ and used this to place a
limit on certain models of quark compositeness.  We use this same ratio
of spectra to place limits on the effects of flavor-universal colorons.

We calculated the leading-order (LO) dijet spectrum 
\begin{eqnarray}
  \kappa &\equiv& {d^3\sigma\over{dM d\eta_1 d\eta_2}} (AB \to 2\,{\rm
    jets}) \\
&=& \sum_{abcd} {x_a x_b M \over {2 \cosh^2{\eta_{cm}}}}
  [f_{a/A}(x_a) f_{b/B}(x_b) + (A \leftrightarrow B, a\neq b)]
  {d\sigma\over d\hat{t}}(ab \to cd) \nonumber
\end{eqnarray}
in terms of the jets' pseudorapidities ($\eta_1$, $\eta_2$; $\eta_{cm}
\equiv 0.5(\eta_1 - \eta_2)$), combined
invariant mass ($M$) and momentum fractions ($x_a, x_b$); the parton
distribution functions ($f_{a/A}, f_{b/B}$); and
the two-body parton scattering cross section
\begin{equation}
{d \sigma\over d\hat{t}}(ab \to cd) = {\pi \alpha_s^2 \over
\hat{s}^2} \Sigma(ab \to cd).
\end{equation}
The leading QCD contributions to $\Sigma(ab \to cd)$ may be found in
\cite{QCDcr} and the contributions from heavy coloron exchange are
given\footnote{In ref. \cite{newint}, the colorons' contributions are
  the terms in equations (4.2-4.5) in that paper which depend on
  coefficient $c_1$.} in \cite{newint}.  Note that we included only
production of gluons and light quarks in our calculations since produced
top quarks would not contribute appreciably to the D\O\ dijet
sample\footnote{The number of top quarks produced is relatively small
  and a decaying top quark does not generally resemble a single
  high-E$_T$ jet.}

Taking a series of different values for the coloron interaction strength
$M_c/\cot\theta$, we then determined the ratio
$\kappa(\vert\eta_{jet}\vert < 0.5) /
\kappa(0.5<\vert\eta_{jet}\vert<1.0)$ at values of $M$ corresponding to
the weighted center of each mass bin\footnote{As a check, we
  re-calculated the ratio for the very wide highest mass bin by
  integrating over $M$ for 800 GeV $< M <$ 1400 GeV; this yielded the
  same ratio as keeping $M$ fixed at the weighted center value of 873.2
  GeV.} measured by D\O\ in ref.\cite{didata}.  We evaluated the
fractional difference between the dijet spectra for pure QCD and for the
various values of $M_c/\cot\theta$.  To simulate a next-to-leading order
(NLO) prediction of the effects of colorons, we multiplied a NLO QCD
prediction obtained using the {\sc jetrad} program \cite{jetr} by the LO
fractional differences.  The results are illustrated in Figure 1.

\begin{figure}[htb]
\centerline{\epsfig{file=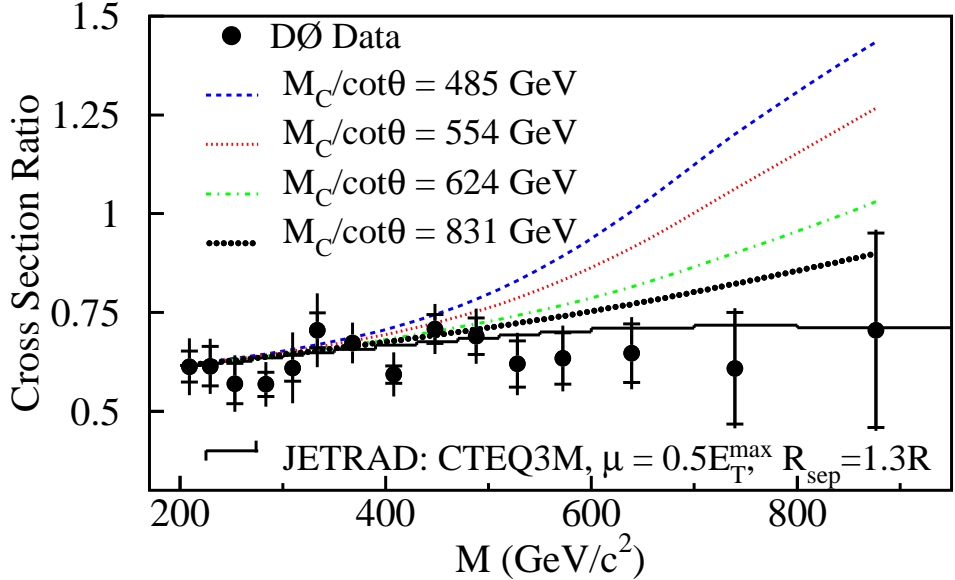, height=8cm}}
\caption[one]{\small The ratio of cross sections, 
$\kappa(\vert\eta_{jet}\vert < 0.5) /
\kappa(0.5<\vert\eta_{jet}\vert<1.0)$, as measured by 
D\O ~\cite{didata} compared to theory for different values of 
\mbox{$M_{c} / \cot{\theta}$} (see text for details on how the
coloron distributions are calculated). The error bars show the
statistical and systematic uncertainties of the data added in
quadrature, and the crossbar shows the size of the statistical error.}
\end{figure}

We extract a limit on the coloron interaction strength from the D\O\
data \cite{didata} using Bayesian techniques with a Gaussian likelihood
function 
\begin{equation}
P(x) = {1\over {{\rm det} S\, 2\pi^2}}\  \exp\left(-{1\over 2} [d -
f(x)]^T\, S^{-1}\, [d - f(x)] \right)
\end{equation}
where $d$ is the vector of data points for the different mass bins,
$f(x)$ is the vector of theory points for the different masses at
different values of $x \equiv 1/\Lambda^n$, and $S$ is the covariance
matrix.  Motivated by the form of $L_{eff}$ (eqn. (\ref{lamta})), the
prior probability is assumed to be flat when $x = 1/\Lambda^2$.  Since
the ratio of spectra at NLO is sensitive to the choice of $\mu$ and
parton distribution function, each possible choice is treated as a
different theory.  The 95\% confidence limit (CL) on $\Lambda$ is
calculated by requiring that
\begin{equation}
Q(x) = \int^x_0 P(x) dx = 0.95\, Q(\infty)\ \ .
\end{equation}
The limit in $x$ is then transformed back into a limit on $\Lambda$ and
thence into a limit on $M_c/\cot\theta$.

The most conservative lower bound we obtain at the 95\% CL is
$M_c/\cot\theta > 837$ GeV for the CTEQ3M pdf and $\mu = \Etmax$
(where \Etmax\ is the maximum jet $E_T$ in the event) as illustrated
in Figure 2.  This limit is incompatible with the suggestion of a
coloron interaction strength of order 700 GeV
\cite{newint} in earlier measurements of the high $E_T$ jet inclusive
cross-section \cite{CDFexc}.

\begin{figure}[htb]
\centerline{\epsfig{file=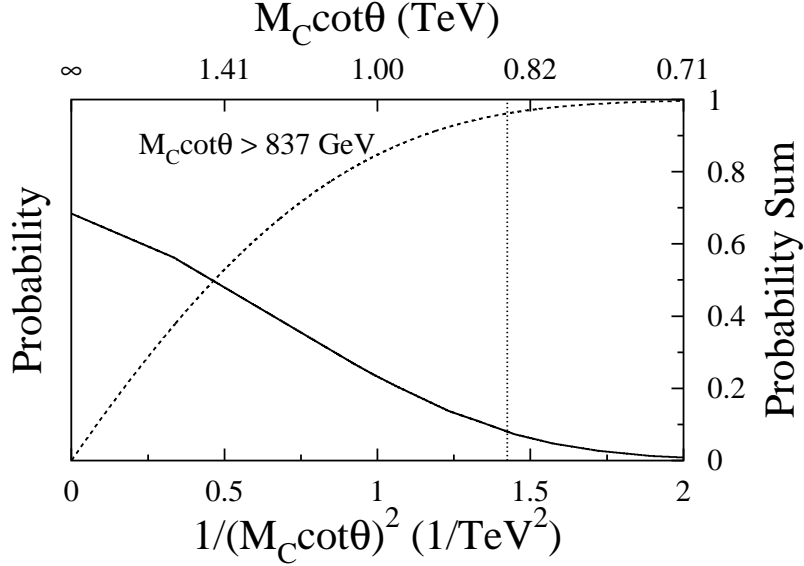, height=8cm}}
\caption[two]{\small The probability distribution (solid curve) 
  $Q(x)/Q(\infty)$ for the theoretical ({\sc jetrad}) prediction with
  $\mu = 1.0\times\Etmax$. The dashed curve shows the integral of the
  probability distribution and the dotted line shows the 95$\%$ CL on
  $M_c/\cot\theta$, i.e., 837 GeV.}
\end{figure}

%%%%%%%%%%%%%%
\section{Discussion}
\label{sec:discussion}
\setcounter{equation}{0}

Our limit places a new exclusion bound in the $M_c$ vs. $\cot\theta$
parameter space of the flavor-universal coloron model.  As shown in
Figure 3, this improves on the recent D\O\ limit based on the dijet
angular distribution \cite{anglim}.  Note that the region at $\cot\theta
\approx 1$ where our limit appears to provide a direct lower bound on
$M_c$ has already been excluded by CDF's search for new particles
decaying to dijets \cite{CDFdij}.  This is fortunate, because our limit
actually becomes less reliable here: the condition $M_c >
\sqrt{\hat{s}}$, under which (\ref{conta}) is a reasonable approximation
to coloron exchange, would no longer hold for the highest-energy data
point.  An updated search for new resonances decaying to
dijets would be a useful complement to the bounds we report here.

\begin{figure}[htb]
\centerline{\epsfig{file=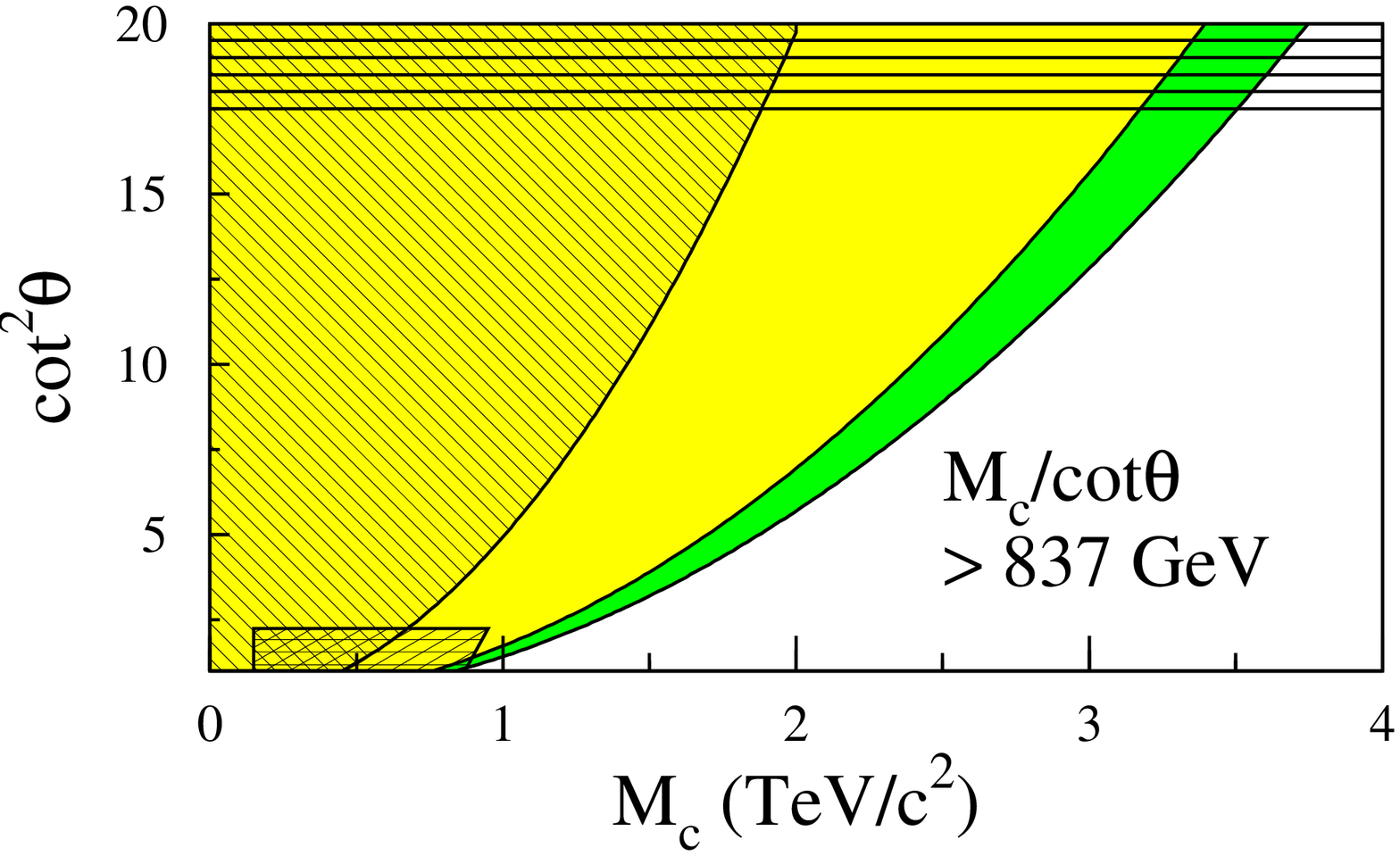, height=8cm}}
\caption[three]{\small Limits on the coloron parameter space: coloron
  mass  $M_c$ vs.
 mixing parameter $\cot{\theta}$. The dark shaded region shows the
 95$\%$ CL exclusion region for the D\O\ measurement of the ratio of
 cross sections (\mbox{$M_{c} / \cot{\theta} > 837$ GeV/c$^2$}). The
 lightly shaded region shows the area excluded by the D\O\ dijet
 angular distribution~\cite{anglim} (\mbox{$M_{c} / \cot{\theta} >
 759$ GeV/c$^2$}). The horizontally hatched region at large
 $\cot{\theta}$ is not allowed in this phase of the
 model~\protect{\cite{newint,colphen}}. The diagonally hatched region
 is excluded by the value of $\rho$ (\mbox{$M_{c} / \cot{\theta} >
 450$ GeV/c$^2$})~\cite{newint}. The cross--hatched region at low
$\cot^2\theta$ is excluded
 by the CDF search for new particles decaying to
 dijets~\protect{\cite{rmh}}.}
\end{figure}

In the context of the dynamical electroweak symmetry breaking model of
\cite{zpcol} in which flavor-universal colorons help produce the mass of
the top quark, the value of $\cot\theta$ is approximately 4.  In
other words, the interaction strength is near its upper limit for
the Higgs phase of the model \cite{newint,colphen}.  Our bound implies that
$M_c > 3.4$ TeV in such models, placing them at the upper right of the
allowed region in Figure 3.

Our findings also set a limit on a broader array of new strong
interaction physics.  Writing eqn.(\ref{conta}) in the more conventional
form for compositeness studies (\ref{lamta}) shows that our limit is
equivalent to a lower bound $\Lambda_{V8} > 2.4$ TeV on the scale of new
color-octet vectorial current-current interactions.  Such interactions
could arise from quark compositeness or from non-standard gluon
interactions (e.g.  gluon compositeness) \cite{chosimm}.

Finally, we have used the same methods to set a limit on a color-octet
axial-vector current-current interaction among quarks.  This has the
form
\begin{equation}
{\cal L}_{eff} = -
{4\pi \over {2! \Lambda_{A8}^2}} J_{5\mu}^a J_5^{\mu a}\ .
\label{axta}
\end{equation}
where $J_{5_\mu}^a \equiv \sum_f {\bar q}_f \gamma_\mu \gamma_5
\frac{\lambda^a}{2}q_f $.  The contributions of this contact interaction
to parton scattering are given\footnote{The relevant terms are those
in equations (4.2-4.5) in that paper which depend on coefficient $c_2$.} in
\cite{newint}.  The most conservative bound we find is $\Lambda_{A8} >
2.1$ TeV at 95$\%$ CL.  Note that we cannot self-consistently
interpret this as providing a lower bound on the mass of an
axigluon \cite{axig, axiphenom} whose exchange underlies the contact
interaction (\ref{axta}).  The relation $\Lambda_{A8} \sqrt{\alpha_s}
= M_{axigluon}$ (since $\cot\theta = 1$ for axigluons) shows that the
mass of the supposed axigluon would not satisfy the condition
$M_{axigluon} >
\sqrt{\hat{s}}$.  Instead, our limit should be interpreted as bounding
the scale of contact interactions arising in models of quark compositeness.

%%%%%%%%%%%%%%
\bigskip \centerline{\bf Acknowledgments} \vspace{12pt} We thank the
D\O\ collaboration for making the measurements that made this work
possible. We also thank R.S. Chivukula for comments on the manuscript.
E.H.S. acknowledges the support of the NSF Faculty Early Career
Development (CAREER) program and the DOE Outstanding Junior
Investigator program. I.B. thanks the D\O\ Collaboration for their
support and contributions to this work and, W.T. Giele,
E.W.N. Glover, and D.A. Kosower for help with {\sc jetrad}. {\em This
work was supported in part by the National Science Foundation under
grant PHY-95-1249 and by the Department of Energy under grant
DE-FG02-91ER40676 with Boston University and grant DE-FG02-91ER40684
at Northwestern University.}

%%%%%%%%%%%%%% bibliography
%\newpage


\begin{thebibliography}{99}
\frenchspacing

\bibitem{newint} ``New Strong Interactions at the Tevatron?'', R.S.
Chivukula, A.G. Cohen, and E.H. Simmons, hep-ph/9603311, to appear in
Physics Letters B (1996).

\bibitem{CDFexc} ``Inclusive Jet Cross Section in $\bar p p$
Collisions at $\sqrt{s} = 1.8$ TeV'', CDF Collaboration, F.~Abe {\it
et al.}, FERMILAB-PUB-96/020-E, hep-ex/9601008.

\bibitem{topglu} C.T. Hill Phys. Lett. {\bf B266} (1991) 419;
C.T. Hill and S.J. Parke, Phys. Rev. {\bf D49} (1994) 4454.

\bibitem{zpcol} M.B.~Popovic and E.H.~Simmons, `A Heavy Top Quark from
  Flavor-Universal Colorons'. To appear in Physical
  Review D. hep-ph/9806287.

\bibitem{kdlr} K.D.~Lane, Phys. Lett. {\bf B433} (1998) 96.
hep-ph/9805254 

\bibitem{tc2} C.T.~Hill, Phys. Lett. {\bf B 345} (1995) 483. hep-ph/9411426.

\bibitem{rmh} R.M. Harris, private communication.  See preliminary CDF
results on the World Wide Web at
http://www-cdf.fnal.gov/physics/new/qcd/qcd\_plots
/twojet/public/dijet\_new\_physics.html\ \ \ .   

\bibitem{colphen} E.H.~Simmons, Phys. Rev. {\bf D55} (1997) 1678.
  hep-ph/9608269; E.H.~Simmons, hep-ph/9701282 and hep-ph/9608349.

\bibitem{anglim} B.~Abbott et al. (D0 Collaboration), Presented at the
  XXIX International Conference on High Energy Physics - ICHEP98, July
  23-29, 1998, Vancouver, B.C., Canada. FERMILAB-Conf-98/279-E.
  hep-ex/9809009.

\bibitem{didata} B.~Abbott et al. (D0 Collaboration),
  Fermilab-Pub-98/220-E, hep-ex/9807014.

\bibitem{elp}  E.~Eichten, K.~Lane, and M.~E.~Peskin,
Phys. Rev. Lett. {\bf 50} (1983) 811.


\bibitem{QCDcr} B.L. Combridge, J. Kripfganz and J. Ranft, Phys.
Lett. {\bf B70} (1977) 234; J.F. Owens and E. Reya, Phys. Rev. {\bf
D18} (1978) 1501.

\bibitem{CTEQ} H.L.~Lai et al. (CTEQ Collaboration), Phys. Rev. {\bf
    D51} (1995) 4763.

\bibitem{jetr} W.T.~Giele, E.W.N.~Glover and
  D.A.~Kosower, Nucl. Phys. {\bf B403} (1993) 633. hep-ph/9302225.

\bibitem{bayes} H.~Jeffreys, {\it Theory of
    Probability} (Clarendon Press, Oxford, 1939, revised 1988), p 94.

\bibitem{CDFdij} CDF Collaboration (F. Abe et al.)
Phys. Rev. Lett. {\bf 74} (1995) 3538.  hep-ex/9501001.


\bibitem{chosimm} E.~H.~Simmons, Phys. Lett. {\bf B226} (1989) 132 and
Phys. Lett. {\bf B246} (1990) 471; P.~Cho and E.~H.~Simmons
Phys. Lett. {\bf B323} (1994) 401 and Phys. Rev. {\bf D51} (1995) 2360.



\bibitem{axig} J. Pati and A. Salam, Phys. Lett. {\bf 58B} (1975) 333;
J. Preskill, Nucl. Phys. {\bf B177} (1981) 21; L. Hall and A. Nelson,
Phys. Lett. {\bf 153B} (1985) 430; P.H. Frampton and S.L. Glashow,
Phys. Lett. {\bf B190} (1987) 157 and Phys. Rev. Lett. {\bf 58} (1987)
2168.

\bibitem{axiphenom} J. Bagger, C. Schmidt, and S. King, Phys. Rev. 
{\bf D37} (1988) 1188. 


\bibitem{bbdij} ``Search for New Particles Decaying to dijets, $b\bar
b$ and $t\bar t$ at CDF'' The CDF Collaboration (R.M. Harris for the
collaboration),  Fermilab-Conf-95/152-E, June 1995.  Published in
Proceedings of the 10th Topical Workshop on Proton-Antiproton Collider
Physics, Fermilab, May 9-13, 1995. hep-ex/9506008. 

\end{thebibliography}
\end{document}